\documentclass{SCIS2018}

\graphicspath{{./grayfigs/}}

\usepackage{xcolor}

\usepackage{multirow}
\usepackage{slashbox}
\usepackage{makecell}
\usepackage{epstopdf}

\begin{document}
\ArticleType{RESEARCH PAPER}{}
\Year{2018}
\Month{January}
\Vol{61}
\No{1}
\DOI{10.1007/s11432-016-0501-5}
\ArtNo{012104}
\ReceiveDate{July 24, 2016}
\ReviseDate{October 21, 2016}
\AcceptDate{December 1, 2016}
\OnlineDate{}

\title{Non-uniform EWMA-PCA based cache size allocation scheme in Named Data Networks}{Non-uniform EWMA-PCA based cache size allocation scheme in Named Data Networks}

\author{Narges MEHRAN}{n.mehran@eng.ui.ac.ir}
\author{Naser MOVAHHEDINIA}{}

\address{Department of Computer Architecture, Faculty of Computer Engineering, University of Isfahan, Isfahan {\rm 81746}, Iran}

\AuthorMark{Mehran N}

\AuthorCitation{Mehran N, Movahhedinia N}

\abstract{As a data-centric cache-enabled architecture, Named Data Networking (NDN) is considered to be an appropriate alternative to the current host-centric IP-based Internet infrastructure. Leveraging in-network caching, name-based routing, and receiver-driven sessions, NDN can greatly enhance the way Internet resources are being used. A critical issue in NDN is the procedure of cache allocation and management. Our main contribution in this research is the analysis of memory requirements to allocate suitable Content-Store size to NDN routers, with respect to combined impacts of long-term centrality-based metric and Exponential Weighted Moving Average (EWMA) of short-term parameters such as users behaviours and outgoing traffic. To determine correlations in such large data sets, data mining methods can prove valuable to researchers. In this paper, we apply a data-fusion approach, namely Principal Component Analysis (PCA), to discover relations from short- and long-term parameters of the router. The output of PCA, exploited to mine out raw data sets, is used to allocate a proper cache size to the router. Evaluation results show an increase in the hit ratio of Content-Stores in sources, and NDN routers. Moreover, for the proposed cache size allocation scheme, the number of unsatisfied and pending Interests in NDN routers is smaller than the Degree-Centrality cache size scheme.}


\maketitle

\section{Introduction}

According to recent predictions \cite{1}, IP multimedia traffic (e.g., TV, Video on Demand (VoD), Internet, and P2P) will continue to be in range of 80 to 90 percent of global traffic by 2019. The growing demand for different sort of contents has led to development of Internet architecture based on Named Data Objects (NDOs), such as documents, videos, or other pieces of information. As a prologue, the common existing techniques use peer-to-peer (P2P) overlays or server-based content distribution networks (CDNs). In these networks, \textit{Consumer} is redirected to the chunks of a copy of the desired content, acquired by the traditional host-to-host networks. Well-known models of P2P networks are Skype  and Bit Torrent  in which peers are responsible for delivering contents like videos or music to other peers. However, being delivered in overlay layers, the requested contents are not being aware of by the routers. Moreover, CDNs requires dedicated-setup for various contents, and optimal selection of peers in P2P networks imposes heavy loads on links \cite{2}. These efforts helped to propose an architecture for re-designing the Future Internet, called Information-Centric Network (ICN) \cite{3,4}, which has in-network caching to efficiently deliver sub-objects (or chunks) to users. Content-Centric Networking project \cite{5} (recently known as Named Data Networking (NDN) \cite{6}) is one of ICN architectures proposed by PARC (Palo Alto Research Center, formerly Xerox PARC) in 2009. The goal of designing this network is to perform management, routing and resource allocation according to the ID considered for the content, instead of the host ID. Indeed, ICN chooses to search for content by asking, ``what's in it?", rather than host-centric approach of ``where's the location of holder?"

	As in Figure 1, every NDN router has three data structures to handle the Interest and Data message forwarding: a Content Store (\textit{CS}), which may be called ``router's cache", or simply ``cache", Pending Interest Table (\textit{PIT}), and Forwarding Information Base (FIB). \textit{CS}, as the name implies, stores the chunks of incoming Data for satisfying possible future Interest (requested by a user) in its storage device, simply named cache. \textit{PIT} role is to store requests not satisfied by \textit{CS} and passed to next router. Finally, FIB role is to forward Interests to the proper face of an NDN router. On the way that Data message is forwarding back to its destination, if an entry in \textit{PIT} is matched, the Data chunk is immediately forwarded to the user.

	In this article, our focus is on studying the importance degree of router in an NDN to improve the performance of a named based cache-enabled network, and to reduce the end-to-end delay and bandwidth consumption as well. To improve an NDN router functionality, the estimation of cost and energy consumption should be considered \cite{7}. Therefore, the issue of changing the structure of nodes is a challenging problem. Trade-off between the high-speed DRAMs and lower-cost SSDs leads to a decrease in costly memory consumption by wisely reducing the speed. Hence, we use our proposed metric related to each node as a base to non-uniformly distribute the amount of high-speed but costly cache among content routers. Consequently, two parameters, the number of Interests satisfied in each router and the number of Pending Interests (PI), are collected to recognize a router's importance in honoring unsatisfied users' Interests. Considering the short-term metrics of outgoing traffic distribution, and user's incoming pending requests, as well as the long-term metric of topology-related Betweenness-centrality of every content router, a novel method for cache size allocation is derived.

	As the former two short-term parameters may change frequently, it is logical to consider our observations in a time interval and calculate the Exponentially Weighted Moving Average. Next, the results are combined with Betweenness parameter of routers to use topological information to size the cache of routers based on different centrality metrics. As reported in \cite{8}, the attainable gain with non-uniform cache sizing is limited, and a centrality-based metric is not sufficient to find the caching performance of routers; however, as the way nodes are linked to one another, it cannot be ignored in getting the importance of routers. Therefore, this paper aims to show that the use of the long-term static metric in conjunction with the short-term dynamic metrics provide the benefits of both methods. PCA-based analysis, as a computationally low-cost approach, reveals relations in parameters that may not be clear. As that combination may increase the size of input data, the important features are extracted by a PCA method to denoise the gathered data and reduce the computational complexity of the algorithm.
	
	In the following section, the related work presented on NDN and specifically their storage devices are briefed. In Section 3, our main contribution is explained. In Section 4, the evaluations for the proposed method are presented, and in Section 5, the paper is concluded.
	
\begin{figure}[!t]
\centering
\includegraphics[scale=.55]{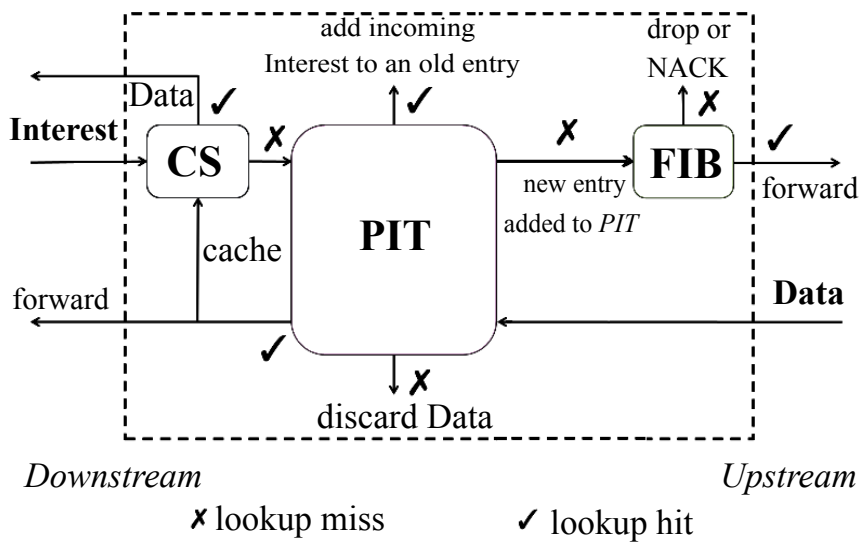}
\caption{Interest and data processing in an NDN router \cite{6}.}
\label{fig1}
\end{figure}

\section{Related work}

$ProbCache$ \cite{9} assesses the caching capability of a path and caches contents probabilistically in order to leave caching space for other data chunks, and acceptably distribute contents in caches along the path from the \textit{Producers} to the \textit{Consumers}. By approximating the distance, that content travels from \textit{Producer} to \textit{Consumer}. Measuring the content capability of a path, NDN router caches the incoming content. Both uniform and non-uniform cache sizes have been adjusted to fit in both ecosystems.

	Authors in \cite{10} design a ``cost-aware cache decision policy", that jointly consider content popularity and price of links through which the contents have to be retrieved. With the help of cost-aware modular decision policy designed, any classic decision policy can be employed. Cost-aware policies allow caching just the contents that are costly for the operator because they reduce the load on expensive and cheap links, while increasing the load on free links. According to their evaluations, by tuning the trade-off between popularity and price, the cost-aware performance is classified into mostly popularity-driven, balanced, and mostly cost-driven categories. A larger gain is achieved in balanced mode, since designing a better composition of hit ratio and cost fraction.

	Cache size allocation schemes have been the subject of other studies, which are closer to our contribution in this paper such as \cite{8,11,12,13}. The centrality metrics of routers have been considered in \cite{8} and been found that the achievable gain with the non-uniform cache size allocation method is not significantly better in comparison with uniform cache size allocation scheme. It has been noticed that the Degree-Centrality metrics improve cache hit for all topologies except Abilene topology where the degree metric does not vary significantly across the nodes.

	Authors in \cite{11} have introduced a novel cache size optimization scheme and classify cache routers based on their roles in content delivery. The information about users' behavior and network traffic distribution are collected by the number of received requests, requests served and content replacements. Their scheme has showed better performance than the graph-related schemes as the important routers have been allocated 80\% of the total cache resources.

	In \cite{12}, a scheme has been presented to give higher cache resource to applications with high performance requirement. In order to assign priority to applications, a manifold learning method has been introduced to perform data mining. The presented evaluations show that the hit ratio of applications such as HTTP web and file sharing is more than Web media and User Generated Contents (such as blogs, chats, tweets). As claimed, the differences are from popularity and average content size of the applications.

	A new metric has been presented in \cite{13}, named Request Influence Degree (RID), to reflect the importance of every router along the content delivery path by considering incoming Interest and outgoing Data messages. The number of all requests received by a router from others has been calculated and multiplied by a weight. The weight is computed from miss rates of caches along the path to show the number of requests that should be sent by a \textit{Consumer} to reach an in-network router. An Interest message arrived at a special router has this information of past misses. However, as the current router needs to have knowledge about past routers on the path, the message-passing among routers is increased significantly.
	
	The short-term and the long-term parameters presented in this article, are inspired from Xu et al. \cite{11}, and the work presented by Rossi and Rossini \cite{8}.

\section{Proposed scheme}

The problem investigated in this article is how to allocate cache size non-uniformly; therefore, we present a scheme by working on the content-store structure of NDN routers. At first, a uniform cache size is allocated to all of NDN routers. After collecting short-term dynamic parameters (satisfied Interests and users' pending Interests in a router), analysing and combining with the long-term static parameter, suitable cache sizes are estimated and allocated to different NDN routers. Furthermore, to consider the impact of network topology, which is a constant parameter, the relative information is combined with the mentioned parameters. An important long-term centrality metric is Betweenness, because it shows how many times a router is on the shortest path between every \textit{Producer}-\textit{Consumer} way. In NDN architecture, it may be shown that this router by possessing a more cache space can store more data chunks for future satisfactions. Besides, in NDN, received Interest and Data packets affect cache size. As these data chunks would be routed for the requester based on the \textit{PIT} table, so more interfaces stored in a \textit{PIT} may cause more data chunks expecting to be received. In addition, more number of cache hits actually shows more Interest satisfaction by the cache. This combination is performed by a proper data fusion scheme to denoise and decrease the size of the input data space. The choice of data fusion scheme is explained next.

\subsection{Data-fusion algorithm}

Our objective is the dimensionality-reduction data fusion to remove unimportant attributes and combine the most important information in a low-dimension set. The learning-based dimensionality-reduction data-fusion schemes, like ISOMAP or K-Means, are generally slow and imposing high computational load on processing the input data, especially for large-scale real-network data processing \cite{14}. According to \cite{15}, PCA has this special capability to extract only the most important information from the data by using the first few components estimated from the original data set. Therefore, to reduce the size of information and combine features with different qualities, PCA is selected.

	PCA demonstrates the technique of approximating the \textit{x} variables with linear combinations of \textit{f} uncorrelated factors, where $f\ll x$. As PCA is capable of filtering unrelated data, it can find the hidden correlation among the variables to generate a set of uncorrelated principal components. As illustrated in Figure 2, if the data set \textit{X} is the observations from two features, then PCA is interpreted as a geometrical representation of the two original axes. The new axes are computed from the two largest variances (eigenvalues) of the data set. By choosing just the first principal component, the data can be projected to get the important features of the original information. To this end, first, a covariance matrix is obtained from the data table. Then, the eigenvectors and eigenvalues of the covariance matrix are calculated to perform Eigen-decomposition and to derive the principal components of the data set.

\subsection{Cache allocation scheme}

Considering a total cache size for the network and by processing the output information of PCA, a proper size is allocated to the cache of each router. Denoting $w_i$ as the weight, importance degree of node \textit{i}, obtained by PCA output and $C_\text{total}$ as the overall cache size of routers, the cache size allocated to each router, $c_i$, is obtained proportional to $w_i$ as below:
\begin{equation}
c_i = C_\text{total} \times w_i,~~i \in [1,n], ~~\sum_{i=1}^{n} w_i = 1.
\label{(1)}
\end{equation}

\begin{figure}[!t]
\centering
\includegraphics[scale=.28]{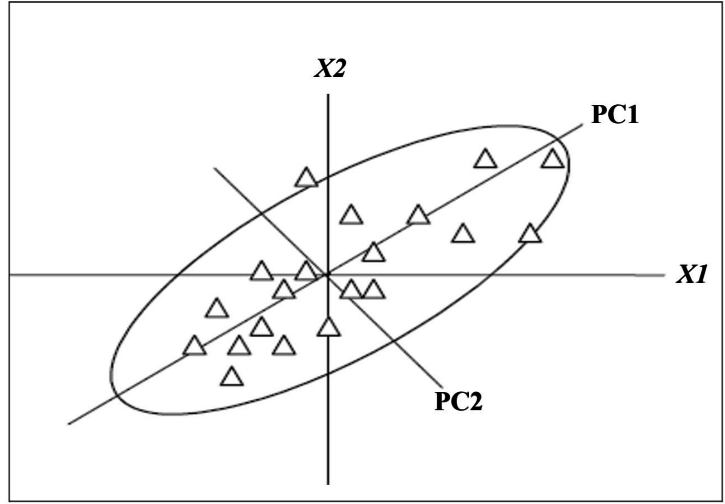}
\caption{Orthogonal transformation of observations from the original feature-based axes to the principal components.}
\label{fig2}
\end{figure}

	In uniform cache size allocation scheme, $w_i$ is the same among all the caches in the network; however, in non-uniform cache size allocation scheme, this ratio differs according to the metrics considered and calculated for each router.
	
\subsection{Metric definition}

Considering NDN as a graph, the routers and the point-to-point links between routers show the nodes and the links in the graph correspondingly. As the links are assumed symmetric and full duplex, the equivalent graph would be unidirectional. To find out how influential a node in a group is, a graph-related long-term parameter, namely Betweenness Centrality (BC), precomputed offline, is used. This metric is an indicator of node centrality in the graph of a network. In order to find this type of centrality, Eqs.~(2)--(4) are adopted \cite{16}.

	Eq.~(2) computes the pair-wise dependencies of two nodes, \textit{s} and \textit{t}, on a special node, \textit{v} \cite{16}:
\begin{equation}
\delta{(s,t|v)} = \frac{\sigma{(s,t|v)}}{\sigma{(s,t)}},
\label{(2)}
\end{equation}
where \textit{$\sigma(s,t)$} shows the number of different shortest paths from node $s$ to $t$. In other words, the pair-wise dependency shows the fraction of all the number of geodesics between pair of nodes, \textit{$\sigma(s,t)$}, that include node $v$, \textit{$\sigma(s,t|v)$}.

	By aggregating these computations, we can have \cite{16}:
\begin{equation}
{\delta{(s|v)}} = \sum_{t \in V} \delta{(s,t|v)}.
\label{(3)}
\end{equation}

	Then, for all $s, v \in V$, we can compute the one-sided dependency of $s$ on vertices one edge farther away \cite{16}:
\begin{equation}
{\delta{(s|v)}} = \sum_{{w:\ } \left\lbrace\substack{ {(v,w) \in E \ \textrm{and} } \\ {\text{dist}(s,w) = \text{dist}(s,v) + 1}} \right\rbrace
   }
  {{\frac{\sigma{(s,v)}}{\sigma{(s,w)}}}{.}{(1 + {\delta{(s|w)}})}}.
\label{(4)}
\end{equation}	

	If a node can reach other nodes on a short path or it is on a significant number of shortest paths between other nodes, it can be concluded that the centrality of this node is high. This metric is used in \cite{17} where only the NDN routers with higher values of BC parameter cache the data chunks because they are more important than other nodes and are more likely to get the cache hits.
	
	The next step is to collect information containing the number of pending Interests and the number of data chunks going from the current node. These two metrics are sampled at the routers. In order to smooth out the fluctuations of the obtained samples, a type of Weighted Moving Average (WMA) procedure is applied. In this type of WMA, called Exponential WMA (EWMA) \cite{18}, the current and the past few samples are weighted exponentially and summed. The sampling frequency should be selected based on the behavior of fluctuations; therefore, 100 samples per second (every 0.01~s) is an almost fair one. In this manner, older observations are assigned smaller weights. Finally, once the weights are assigned, a feature value based on the samples of each dynamic parameter is computed for the router.
	
	\textit{New\textunderscore PI\textunderscore Sample(t)} is considered to be the time series of \textit{PIT} snapshots (the number of pending Interests in \textit{PIT} at different moments) and by denoting \textit{PI\textunderscore Avg(t)} as the output of EWMA, the PI prediction is calculated using the current and the past users' PI samples in an NDN router by (5)--(7):
\begin{equation}
{PI\textunderscore Avg(t)}  = {g \times {New\textunderscore PI\textunderscore Sample(t)}} + {(1-g) \times {PI\textunderscore Avg(t-1)}},
\label{(5)}
\end{equation}
where $g$ is a constant value ($0 < g < 1$). A smaller ``$g$" assigns smaller weight to current observation, and gives more importance to long-term history. Having long-term correlation between PI samples, a particularly good selection for ``$g$" is $1/2^{4} = 0.125$ which is a reciprocal power of two. Because by multiplying (5) and 1/$g$, it is possible to keep everything integer and have a faster calculation.

	In addition, it is worthy to have a measurement for variability of observations by computing the standard deviation of EWMA output. In other words, for showing the fluctuation of samples from the average value, another Exponentially Weighted Moving Average (Eq.~(6)) is computed from the absolute difference of current sample and EWMA given by  (5). The selected value for $g$' is considered $(1 / 2^{2}) = 0.25$ that increases the sensitivity of new-sample deviation from the EWMA in  (5):
\begin{equation}
{PI\textunderscore Dev(t)}  = PI\textunderscore Dev(t-1) + {g'{.}\left\lbrace{|New\textunderscore PI\textunderscore Sample(t)-PI\textunderscore Avg(t)|\ - \ PI\textunderscore Dev(t-1)}\right\rbrace}.
\label{(6)}
\end{equation}

	Given the values of average and deviation, then the question is how to estimate the number of PIs. This value should not be smaller than average (Eq.~(5)); otherwise cache size will decrease, which is not a fair allocation in our algorithm. It is, therefore, desirable to assign an average value plus some margin. Here, deviation is used as the margin, because it represents the variation of observations from the average value. When samples greatly fluctuate around the average, it is better to select a large margin; on the contrary, in case of small fluctuations, a small margin is preferred. By minimizing estimation error, the proper equation is obtained as follows (Eq.~(7)). In other words, by considering the minimum estimation error between \textit{New\textunderscore PI\textunderscore Sample(t)} and \textit{Estimated\textunderscore PI(t)}, and changing the factor of \textit{PI\textunderscore Dev(t)}, the estimation of the average number of pending Interests is assessed. By employing the key attribute of EWMA method, the last value of \textit{Estimated\textunderscore PI(t)} is selected by a measurement from the past $t$ seconds of gathered PIs.
\begin{equation}
{Estimated\textunderscore PI(t)} = {PI\textunderscore Avg(t)}+(0.1 \times {PI\textunderscore Dev(t)}).
\label{(7)}
\end{equation}

	The same equations can be considered for the number of Interests satisfied by \textit{CS} in current NDN router. If \textit{New\textunderscore HI\textunderscore Sample(t)} (the number of satisfied Interests in \textit{CS} at different moments) is the time series of the outgoing Hit Interests (HIs) retrieved from \textit{CS} of a router at moment \textit{t}, EWMA of the number of hits in a router is predicted as
\begin{equation}
{HI\textunderscore Avg(t)}  = {g \times {New\textunderscore HI\textunderscore Sample(t)}} + {(1-g) \times {HI\textunderscore Avg(t-1)}}.
\label{(8)}
\end{equation}

	According to  (9), another EWMA is computed from subtraction of current sample and the measured EWMA in  (8):
\begin{equation}
{HI\textunderscore Dev(t)}  = HI\textunderscore Dev(t-1) + {g'{.}\left\lbrace{|New\textunderscore HI\textunderscore Sample(t)-HI\textunderscore Avg(t)|\ - \ HI\textunderscore Dev(t-1)}\right\rbrace}.
\label{(9)}
\end{equation}

	Considering the average value (Eq.~(8)) and the EWMA of deviations (Eq.~(9)), estimation for the number of router's HIs is calculated as below:
\begin{equation}
{Estimated\textunderscore HI(t)} = {HI\textunderscore Avg(t)}+(0.1 \times {HI\textunderscore Dev(t)}).
\label{(10)}
\end{equation}

	In the next section, we will describe how to use PCA to allocate a size to every router's cache, based on these time-variant variables and time-invariant one.

\subsection{The proposed cache size allocation scheme}

After estimating the EWMA of outgoing HIs and incoming users' PIs, the PCA data-fusion algorithm is used to combine the dynamic parameters with static Betweenness-centrality parameter. This algorithm finds the most important information in a new subspace and projects data set on the orthogonal uncorrelated principal components of the subspace. In large-scale data sets, it can be seen as $I \times J$ matrix, where $I$ is the number of routers in the network, and $J$ is the number of features collected from every router. This input is fed to PCA algorithm to calculate the importance of the router in NDN. Then, this value is utilized to allocate proper cache size to the router. The PCA steps are explained below:

\begin{itemize}
\item Normalizing the elements of the data set:
\begin{equation}
x\ = \frac{x -\textrm{mean}}{\textrm{std\textunderscore dev}},
\label{(11)}
\end{equation}
where mean is the average and {std\textunderscore dev} is the standard deviation of the data set.

\item Finding the covariance matrix of the normalized data set (it shows how much the two variables change together):
\begin{equation}
{\text{Cov}_{i,j}}\ =\ {\frac{1}{I}} \sum_{q=1}^{I} {X_{q,i}{.}X_{q,j}}.
\label{(12)}
\end{equation}

\item Calculating the eigenvectors of the covariance matrix for each component as
\begin{equation}
{Q_{i}\textrm{Cov}}\ =\ {{\lambda}_{i}\textrm{Cov}},
\label{(13)}
\end{equation}
where \textit{i} is the global ID for the router \textit{i} in NDN.
\end{itemize}

	By setting a desired value for $L$ (not larger than $J$), it is possible to extract just $L$ number of features from the most important features of the data table. In this article, just the first eigenvector is selected, because the first one is made out of the largest possible variance of the data set.

	Then the data set is projected onto this eigenvector; therefore, the data points in the new subspace are obtained by
\begin{equation}
F_{I \times 1}\ =\ X_{I \times 3}\ {.}\ {Q_{3 \times 1}}.
\label{(14)}
\end{equation}

	A flow diagram of PCA-based data-fusion process is depicted in Figure 3. By showing the parameters collected from each router \textit{i} with: $X_1[i]$, $X_2[i]$, $X_3[i]$, and the principal components of the first eigenvector with: $\text{PC}_1$, $\text{PC}_2$ and $\text{PC}_3$ such that $\text{PC}_1+\text{PC}_2+ \text{PC}_3\ =\ 1$, then the fused data for an NDN router is obtained by
\begin{equation}
F_\text{fused}[i]\ =\ X_{1}[i]\text{PC}_{1}+X_{2}[i]\text{PC}_{2}+X_{3}[i]\text{PC}_{3}.
\label{(15)}
\end{equation}

	At the end, it is essential to normalize the elements of vector \textit{F} to have a ratio for calculating the cache of each router. Eq.~(16) is used to compute the weight of an NDN router:
\begin{equation}
w_i = \frac{F_\text{fused}[i]}{\sum_{i=1}^{n} F_\text{fused}[i]},~~i \in [1,n],~~ \sum_{i=1}^{n} w_i = 1.
\label{(16)}
\end{equation}

	Using this value, the percentage of the overall NDN cache size, which should be allocated to that router, is derived. In other words, by considering the total cache size of the network and the quota of each router, as mentioned in  (1), the suitable size for \textit{CS} of the router is decided.

\begin{figure}[!t]
\centering
\includegraphics[scale=.35]{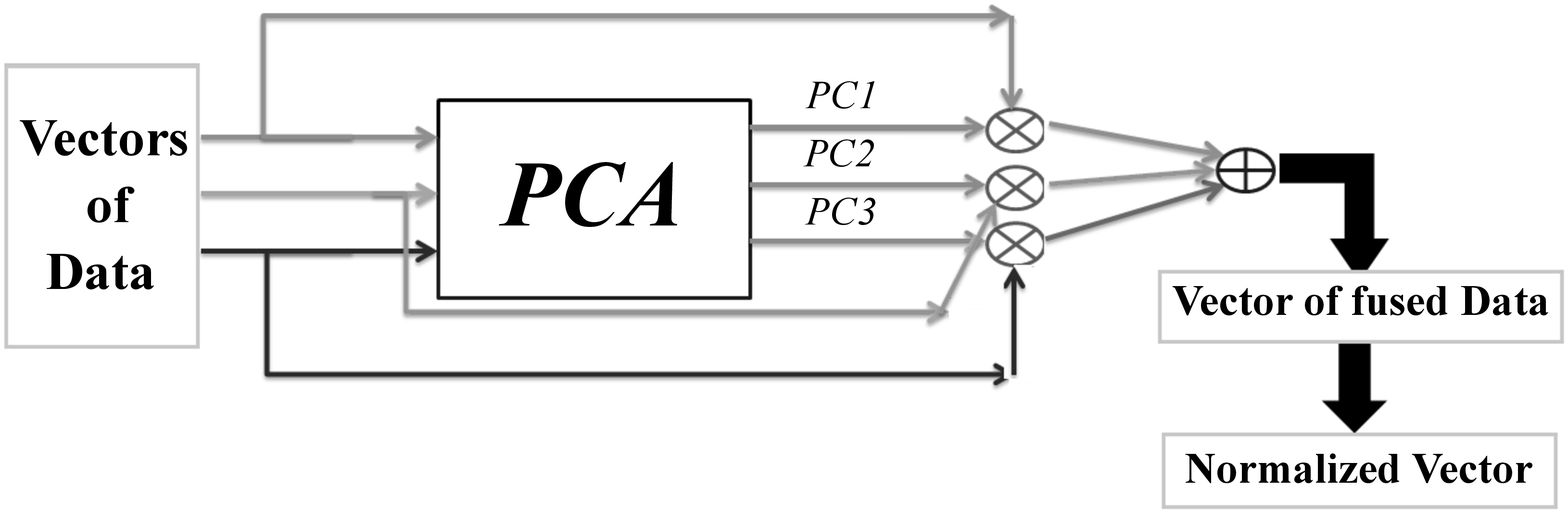}
\caption{The PCA based data fusion diagram.}
\label{fig3}
\end{figure}

	Algorithm 1 presents the proposed scheme. In this algorithm, all the steps are performed offline except the second step, which need the runtime of NDN. As explained, the raw data table, consisting of three aspects of information, is processed by PCA-based data-fusion algorithm \cite{19}. First, by considering a graph for NDN, the centrality metric of every router is computed. Second, by running the NDN network, the short-term parameters are calculated. Third is the place where EWMA of short-term metrics is estimated. In the fourth and fifth steps, the principal components are obtained from the three parameters and the fused data is computed for the router. Then by normalizing the fused data, the importance of routers are figured out. Finally, the algorithm produces the granted cache size based on importance of an NDN router, which needs to be modified to get the new cache storage volume.

\begin{algorithm}[!t]
\footnotesize
\caption{~~Proposed cache size allocation algorithm}
\label{Algorithm}
{\bf { \sc{ProposedCacheSizeAllocation\ ({\textit{Topology of NDN Network}})}}}
\begin{algorithmic}[1]
	\STATE{Calculating Betweenness-centrality of every node offline;}
	\FOR{every NDN router}
		\STATE {Counting the number of PIs in PIT and the number of HIs in current CS;}
	\ENDFOR
	\STATE{Estimating the EWMA of counters in previous step;}
	\STATE{Apply the data-fusion PCA algorithm on vectors of: (Betweenness, EWMA of PIs, EWMA of HIs);}
	\STATE{Using the principal components to calculate the fused data for every router;}
	\STATE{After normalizing the fused-data vector, the importance of every router in NDN is presented;}
	\STATE{By considering an overall cache size for the NDN network, counting every router's cache size.}
\end{algorithmic}
\end{algorithm}

	Assume that output of Step 2 is ready and the additional cost of online runtime is neglected. The algorithm for Betweenness-centrality metric of every router runs in $O(NM)$ time on unweighted networks~\cite{16}, where $M$ is the number of links and $N$ is the number of routers in the NDN network. After gathering the information from Step 2, the load of processing them in Step 3 is about $O(Nt_s)$, where $N$ is the number of NDN core routers and $t_s$ is the number of samples of HIs and PIs. In Step 4, by having just three features from an NDN router, it takes $O(kN)$ to compute the principal components. $k$ is the number of feature. Other steps also takes $O(N)$ to calculate the weights as an importance for every NDN cache. Therefore, the total time complexity of Algorithm 1 is $O(NM+Nt_s+kN)$, which is a polynomial time complexity. In other words, the complexity of this algorithm is dependent on network topology and the number of samples from all the NDN routers' short-term metrics.

\section{Evaluations}

Among ICN architectures, NDN is a more complete paradigm, which includes an SDK (CCNx \cite{5}), and simulation software like ndnSIM \cite{20} and ccnSim \cite{21}. In addition, a new ICN simulator, Icarus, has been developed \cite{22}. This new simulator offers researchers to evaluate their studies in other ICN architectures \cite{23}, but we wanted to mainly focus on an NDN based simulator that can be extended by other TCP/IP protocols with the proposed method in future. Therefore, we evaluate our proposed method by using the open source NS-3 based simulator, ndnSIM-v1 released by the Internet Research Lab of UCLA.

	We test our experiments on a real-network topology called Abilene topology. The Abilene network, in the USA, connects most of the research labs and universities. Figure 4(a) indicates the topology of this network with eleven routers in 2011. Abilene is an advanced backbone network operated by the University Corporation for Advanced Internet Development (UCAID) to support the physical components of the Internet2 project \cite{24}. This topology has been used in various assessments for the CCN researches, like \cite{25}. Our experiments are tested in a topology observed in Figure 4(b). The routers are connected by point-to-point links with a bandwidth of 10 Gbps. Twelve \textit{Consumer} nodes attached to the routers extend the topology. Each \textit{Consumer} may request any of the four different applications running on various \textit{Producer} nodes. The ndn::BestRoute \cite{26} Forwarding Strategy, a module defining how Interest and Data are being forwarded, is selected. This NDN-specific forwarding strategy is able to handle network problems effectively with adaptive forwarding. The \textit{Consumer} applications issue Interests at a configurable frequency, and initiate the data transferring. The chunk size in \cite{27} is considered as packet-level size 1 KB, but in NDN literature, it is expected to be 10 KB \cite{28}.

\begin{figure}[!t]
\centering
\subfloat[]{\includegraphics[scale=.5]{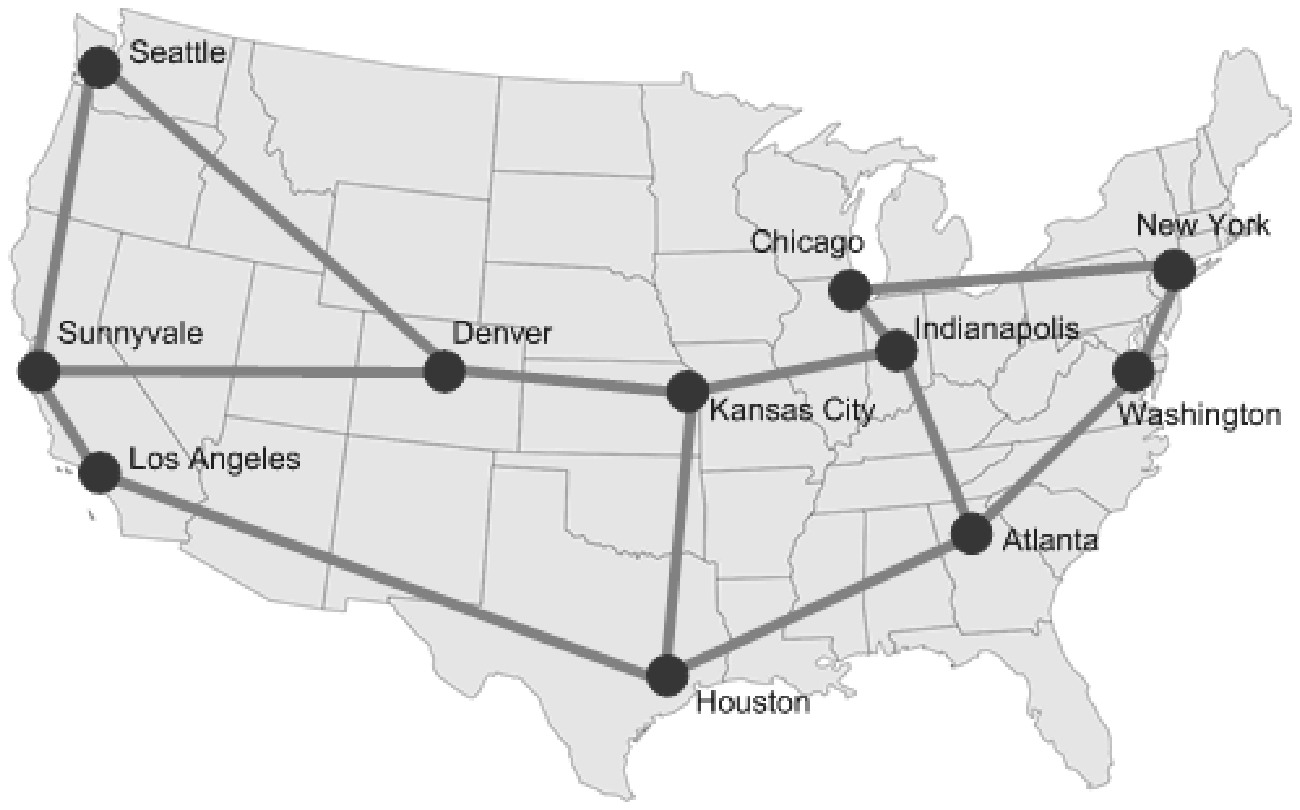}}
\subfloat[]{\includegraphics[scale=.35]{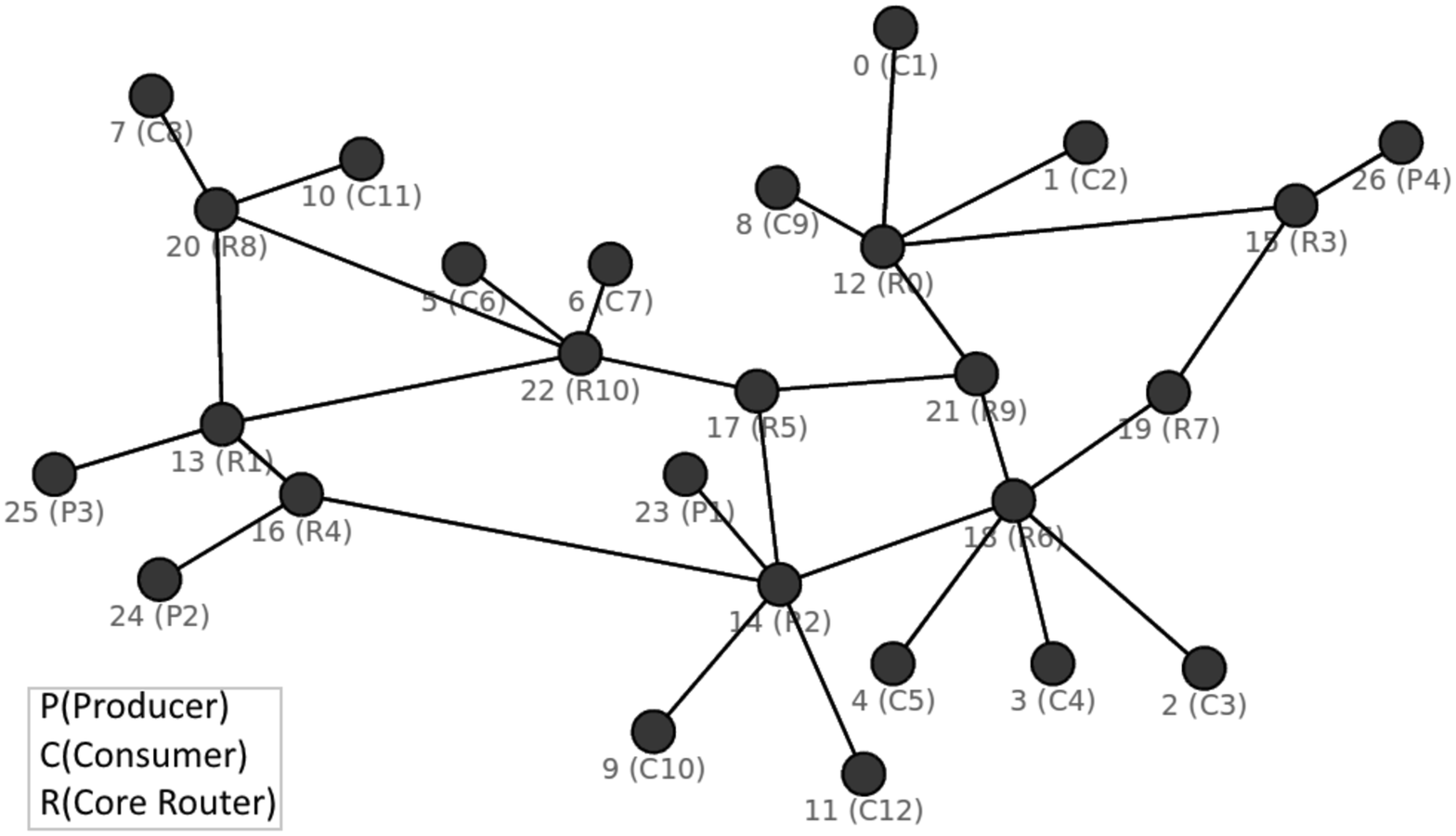}}
\caption{(a) Abilene core topology [24]; (b) simulated topology with Abilene routers.}
\label{fig4}
\end{figure}

	The conditions, under which the simulation is made, are tabulated in Table 1. The cache replacement policy is LRU (Least Recently Used). The total chunk-based in-network cache size of a router is set to $13000$ chunks, corresponding to the ${13 \times {10^{-8}}}$ of the catalog size. It is assumed that every Interest requests a unique chunk of content object at a specific time. This uniqueness is achieved by adding a sequence number to the end of the prefix. Therefore, the \textit{Producer} application is configured to reply to a received interest with the related data chunk.

\begin{table}[!t]
\belowrulesep=0pt 
\aboverulesep=0pt 
\renewcommand{\arraystretch}{1.4} 
\doublerulesep 2.2pt
\caption{System parameters used in the experiments}
\label{tab:title}
\footnotesize
\begin{tabular*}{\textwidth}{@{\hspace*{6pt}}@{\extracolsep{\fill}}ccc@{\hspace*{6pt}}}
\toprule
		 {} & {Parameter} & {Value} \\
	\hline
		\multirow{8}{*}{Network} & {Chunk size}	& {10 KB} \\
			{} & {Cache size}		& 10--128 MB \\
			{} & {Network size}	& {27 {Nodes}} \\
			{} & {Topology}		& {{Abilene, with 11 Nodes}} \\
			{} & {Core network link capacity}	& {{10 Gb/s}} \\
			{} & {Access network link capacity}	& {{1 Gb/s}} \\
			{} & {Rate of interest}	& {{20 Hertz}} \\
			{} & {Simulation time}	& {{500 s}} \\
			{} & {Number of seeds}	& {{10 (generated with normal distribution)}} \\
	\hline
		\multirow{3}{*}{Catalog} & {Catalog size}	&	{$10^7$} \\
			{} &	{File size} & {{100 MB}} \\
			{} &	{(Cache/Catalog) ratio}&{$13\times 10^{-8} $} \\
	\hline		
		\multirow{2}{*}{Popularity}&{Plateau factor in Mandelbrot-Zipf Distribution ($q$)}&{{{0, 5, 50}}} \\
			{}&{Shaping factor in Mandelbrot-Zipf distribution ($s$)}&{\centering	{[0.5, 2.5]}}\\
	\hline		
		\multirow{2}{*}{\parbox{2.4cm}{\centering Forwarding \& caching policy}} &	{Forwarding strategy} & {BestRoute} \\
			{} & {Replacement policy}	& {LRU} \\
\bottomrule
\end{tabular*}
\end{table}

	The content distribution of the network is modeled by Mandelbrot-Zipf distribution as\cite{29}
\begin{equation}
p(i) = \frac{\frac{1}{(i+q)^s}}{{\sum_{j=1}^{|F|} \frac{1}{(j+q)^s}}},~~i \in [1,|F|].
\label{(17)}
\end{equation}
	$p(i)$ denotes probability of retrieving the $i$th content from the number of files $|F|$ in the network. The parameters $q$ and $s$ are fixed for an ISP. $s$ indicates the Mzipf shaping factor in the range of $[0.5, 2.5]$ and as the name implies, it shapes the request arrival pattern. $q$ is the plateau factor, which controls the plateau shape (flattened head) near the lowest ranked contents. If $q$ is equal to zero, then Mzipf degenerates into a Zipf distribution \cite{28}.
	
	By applying this popularity distribution of files, it is observed that from $10^7$ number of files in a catalog such as a large YouTube-like catalog, with $q=5.0$ and $s=0.7$, the requests are forwarded toward 7073017 files, which is above 90\% of users' popular contents \cite{30}. If shaping factor, $s$, is reduced to 2.0, then the number of the most popular files in the catalog is decreased to just six files.

	Therefore, by caching these six popular files, it is possible to decrease the network bandwidth and improve the speed of users' request satisfaction.
	
	The experiments are repeated ten times, all of them are evaluated in 500 s of simulation time, which is about one-day runtime of a real NDN network. Every \textit{Consumer} node issues 20 requests per second. The results, the evaluation, and its metrics are as follows. The normalized values computed for every router, based on the parameters selected in this paper, are shown in Table 2.

\begin{table}[!t]
\belowrulesep=0pt 
\aboverulesep=0pt 
\renewcommand{\arraystretch}{1.4} 
\doublerulesep 2.2pt
\caption{Normalized parameters gathered from the NDN routers}
\label{tab:title}
\footnotesize
\begin{tabular*}{\textwidth}{@{\hspace*{14pt}}@{\extracolsep{\fill}}cccc@{\hspace*{14pt}}}
\toprule
{Parameter}
			&\makebox[3em]{BC} &\makebox[10em]{ EWMA of PIs} &\makebox[10em]{EWMA of HIs} \\			
	\hline
	{NDN Router 0}	& 0.62963791	& 0.57540494	& 0.56401801\\
	{NDN Router 1}	& 0.24072418	& 0.06403307	& 0.33221026\\
	{NDN Router 2}	& 1	& 0.61991863	& 0.96397991\\
	{NDN Router 3}	& 0.06481895	& 0	& 0.96397991\\
	{NDN Router 4}	& 0.31481895	& 0.51107432	& 0.58686915\\
	{NDN Router 5}	& 0.83331471	& 0.00547495	& 0\\
	{NDN Router 6}	& 0.87963791	& 0.81142273	& 0.68312906\\
	{NDN Router 7}	& 0	& 0.17876889	& 0.59078904\\
	{NDN Router 8}	& 0.25	& 0.28850586 &	0\\
	{NDN Router 9}	& 0.78704738	& 0.81981368	& 0\\
	{NDN Router 10}	& 0.75	& 0.99840404	& 0.35154242\\
\bottomrule
\end{tabular*}
\end{table}

	As observed in Figure 5, the number of PIs (averaged on \textit{PIT} tables) follows the Poisson distribution. Indeed, content items and data chunks, are requested by \textit{Consumers} according to a specific workload (i.e., Poisson arrival).

\begin{figure}[!t]
\centering
\includegraphics[scale=.3]{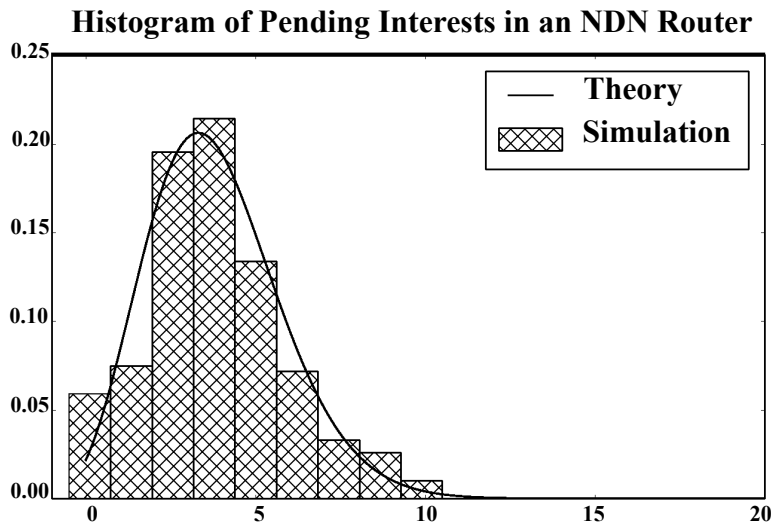}
\caption{Histogram of PIs in average of core routers, ($q=0.0, ~s=2.0$).}
\label{fig5}
\end{figure}

	The load of each category of nodes is shown by the hit ratio, which is the ratio of the requests that are satisfied in every node to the total number of requests (hits plus misses) presented in  (18):
\begin{equation}
HitRatio(t) = \frac{\sum_\text{requests}{Hit(t)}}{\sum_\text{requests}{Hit(t)+Miss(t)}}.
\label{(18)}
\end{equation}
	The average of hit ratios among a set of nodes can be considered to evaluate the performance of all caches; e.g., the set of \textit{Producers}, or the set of NDN routers. As the number of \textit{Producer} nodes in our topology is four, and also the number of routers is eleven, so the average hit ratio is considered for evaluations.
	
	As the performance of the proposed method and previous state of the art overlap in small catalog size, in our evaluations, large catalog size is applied. Therefore, the hit ratios have decreased because different types of chunks need to be stored. Higher traffic and hit ratio of caches requires that the system running the simulations have more powerful processing unit.

	As observed in Figure 6, by assigning a better size to every cache based on the proposed scheme, the router's load is increased. In this way, most of the requests are satisfied before arriving at the \textit{Producer}. The enhancement of the proposed method at the average hit ratio of routers is 20.7\% against Degree-Centrality method and about 17.7\% against the Uniform method. As observed, until the 200 s of the simulation time, the hit-ratio of the methods is close to each other, but the proposed method supersedes the hit ratio of routers' hit ratio in the second half.

	Figure 7 shows that the proposed scheme outperforms the Degree-Centrality allocation scheme and the Uniform scheme in \textit{Producer's} hit ratio. As observed, the proposed scheme increases \textit{Producer's} \textit{CS} hit ratio in comparison with other schemes. The \textit{Producer's} \textit{CS} miss action may be due to the data which is not stored in cache, therefore a miss occurs in \textit{CS} and the \textit{Producer} must generate that content by means of its permanent repository in response to the unsatisfied interest message; however, by increasing the \textit{CS} hit ratio, the permanent repository's load decreases. The average hit ratio in our scheme is 7.9\% better than Degree-Centrality method; the proposed scheme is also 5.5\%, better than Uniform method. The management of in-network caching facilities helps to reduce the references to the repository in application layer. Hence, the \textit{Producer} just needs to store the less popular chunks. The comparison in Table 3 indicates that the chunks of different applications in the proposed method can achieve better number of hits comparing to other methods and improve the overall \textit{Producers'} \textit{CS} hit ratios.

\begin{figure}[!t]
\centering
\begin{minipage}[t]{0.48\textwidth}
\centering
\includegraphics[scale=.35]{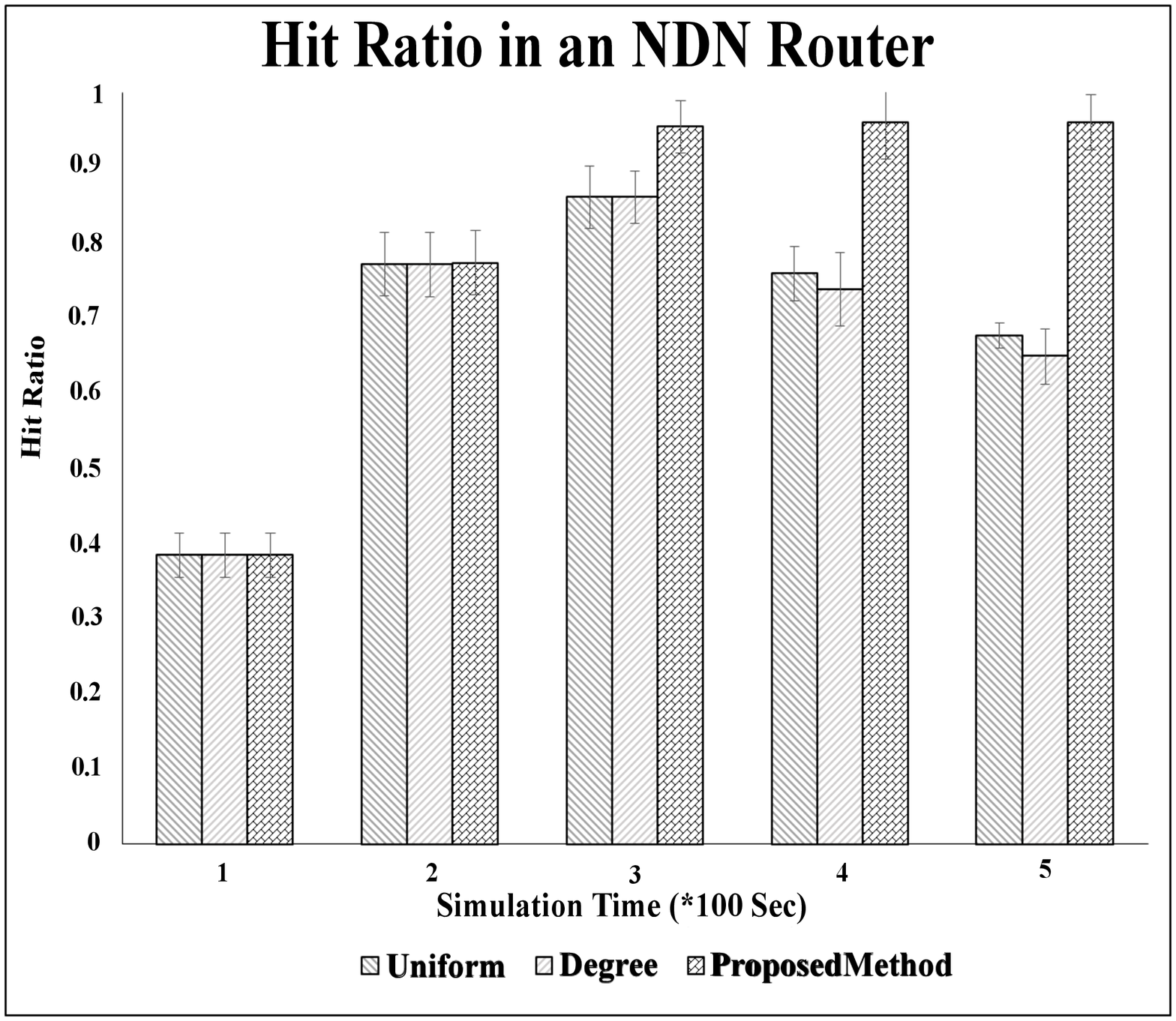}\\
\caption{Router hit ratio  in uniform, Degree-Centrality and proposed method ($q=5.0, ~s=0.7$).}
\label{fig6}
\end{minipage}
\hspace{0.02\textwidth}
\begin{minipage}[t]{0.48\textwidth}
\centering
\includegraphics[scale=.35]{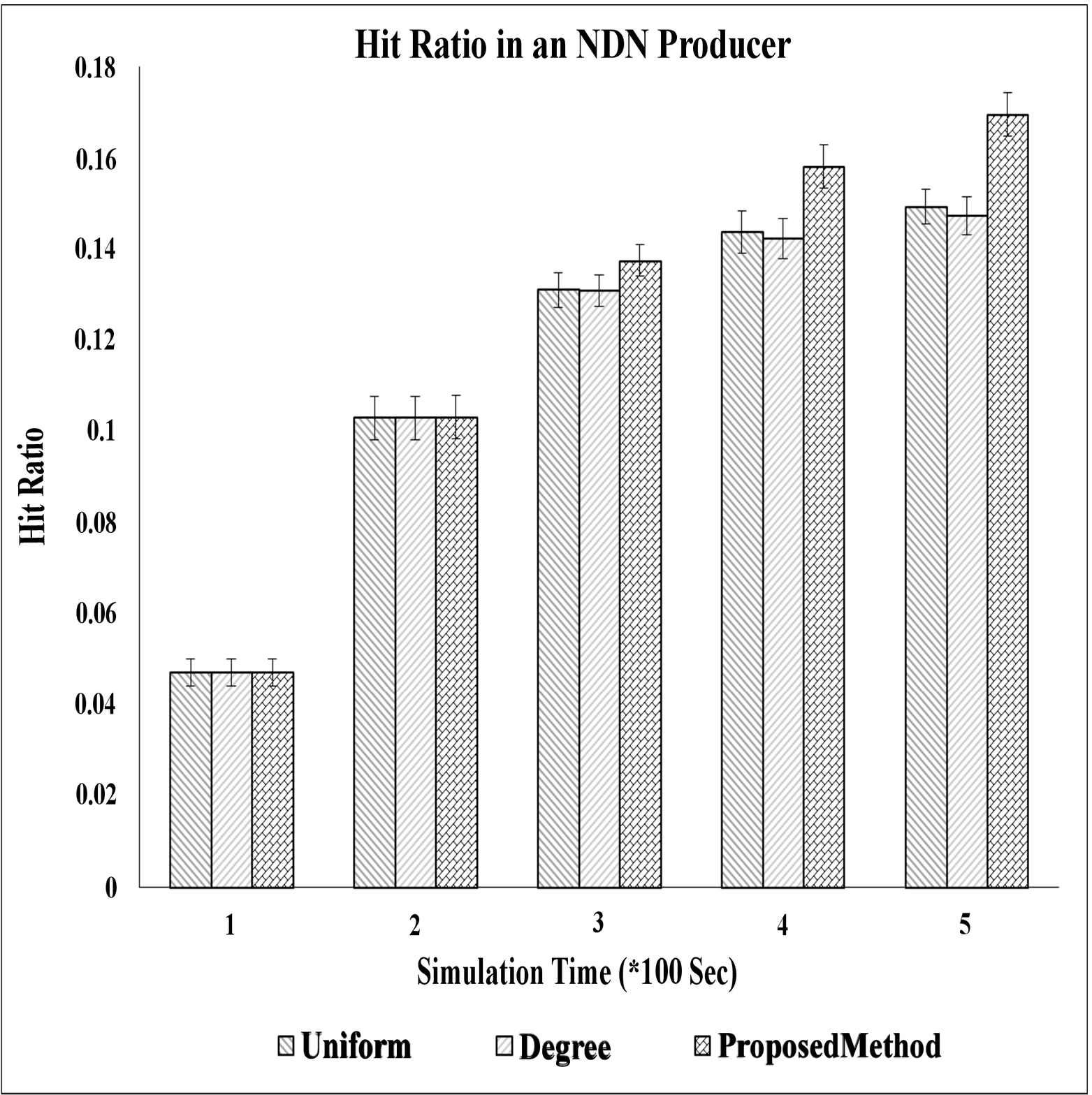}\\
\caption{Producer's content-store  hit ratio  in uniform, Degree-Centrality and proposed method ($q=5.0,~ s=0.7)$.}
\label{fig7}
\end{minipage}
\end{figure}

\begin{table}[!t]
\belowrulesep=0pt 
\aboverulesep=0pt 
\renewcommand{\arraystretch}{1.4} 
\doublerulesep 2.2pt
\caption{Performance of each producer's CS running an application} \label{tab:title}
\footnotesize
\begin{tabular*}{\textwidth}{@{\hspace*{14pt}}@{\extracolsep{\fill}}ccccc@{\hspace*{14pt}}}
\toprule
		& Application 1
			& Application 2
				& Application 3
					& Application 4\\
	\hline
	{ Uniform} & 532 & 898	& 896 & 1329\\
	{ Degree-Centrality} & 528 & 899 & 887	& 1312\\
	{ Proposed method} & 533 & 933	& 899 & 1501\\
\bottomrule
\end{tabular*}
\end{table}

	Figure 8 illustrates the Round Trip Time (RTT) of the transmitted Interest and its delivered Data. As it can be seen, the proposed scheme is slightly different from the Uniform and Degree-Centrality schemes in RTT. The average delay of proposed scheme decreases about 0.15\% in comparison with Uniform method and about 0.12\% in comparison with Degree-Centrality method. This is because of the levels in our topology. In other words, in considered Abilene-based topology, requests are passed from a few routers to reach its data; but if a tree-based topology is considered with more levels of routers, it is expected that the reduction of delay in the cache-size allocation schemes may show their differences.

	The average number of PIs, received and stored in \textit{PITs} of routers, in the proposed method and the Uniform and Degree-Centrality methods are compared in Figure 9. It is observed that by applying the proposed scheme, the number of PIs is reduced about 37\%. It means that by exploiting the proposed cache-size allocation scheme and controlling the Satisfied Interests by increasing the cache size of important nodes, the average number of unsatisfied Interests, in comparison with previous schemes, and consequently the traffic in NDN are decreased. In this graph, it is observed that by increasing the NDN router cache hit, the number of PIs in \textit{PIT} is decreased because the Interests are satisfied earlier.

\begin{figure}[!t]
\centering
\begin{minipage}[t]{0.47\textwidth}
\centering
\includegraphics[scale=.35]{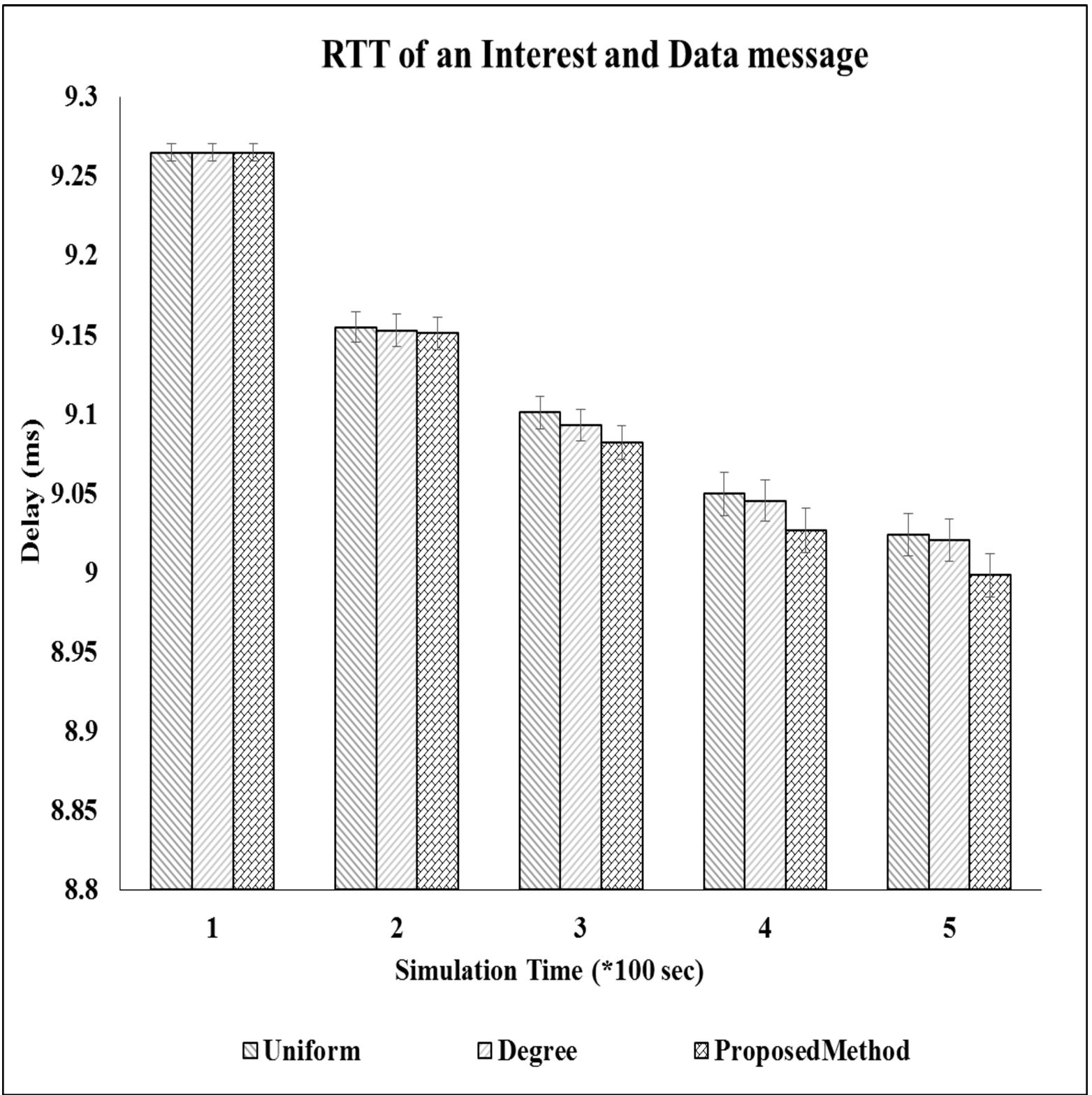}
\caption{Round-trip time of interest-data delay in \mbox{uniform}, Degree-Centrality and proposed method ($q=5.0,~ s=0.7$).}
\label{fig8}
\end{minipage}
\hfill
\begin{minipage}[t]{0.47\textwidth}
\centering
\includegraphics[scale=.35]{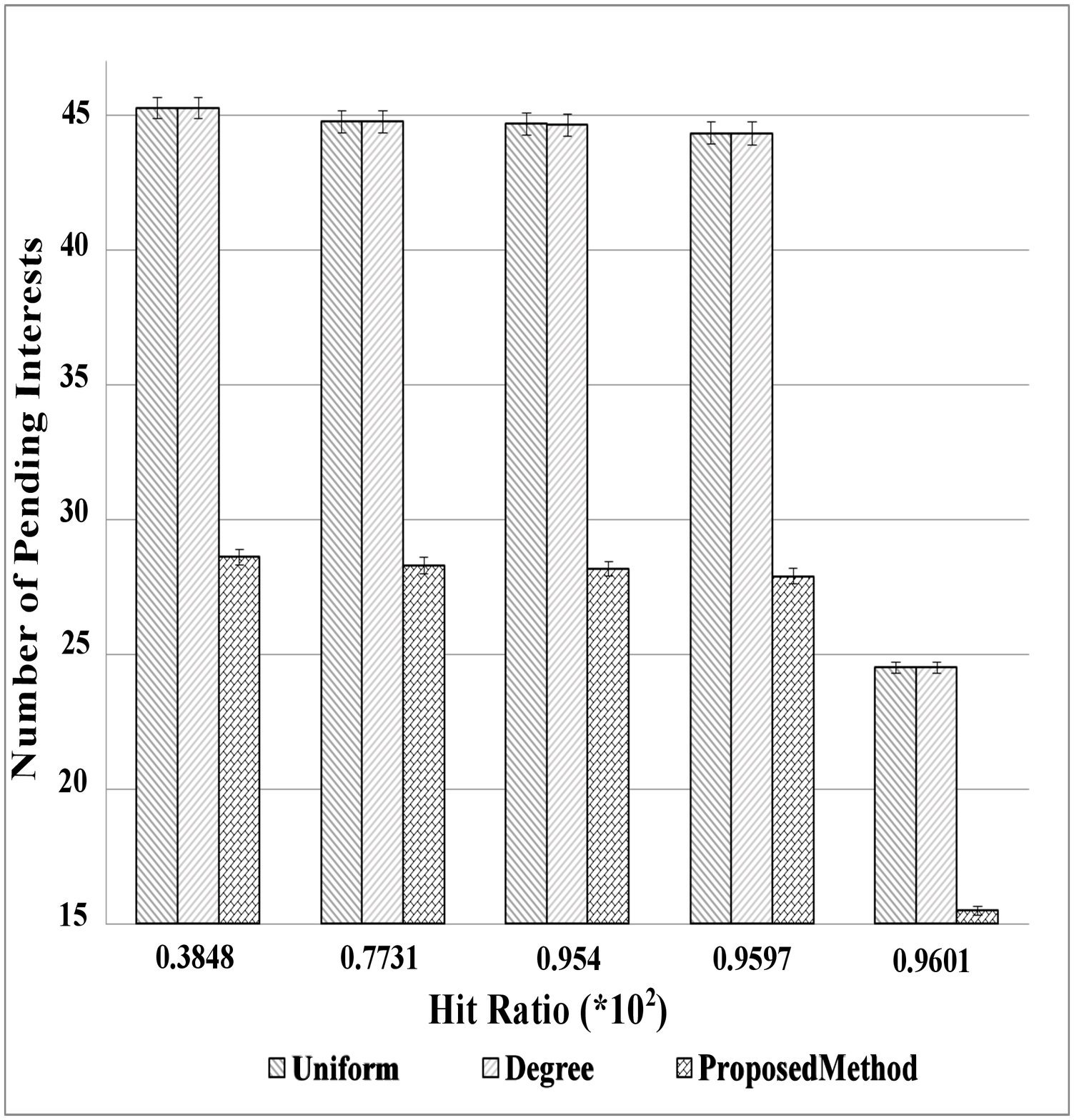}
\caption{Pending interests in PIT, averaged on routers in uniform, Degree-Centrality and proposed method ($q=5.0, ~s=0.7$).}
\label{fig9}
\end{minipage}
\end{figure}

\section{Conclusion}

The router's cache size is one of the essential issues in NDN performance. To assign proper size of cache to each router in this network, a method is proposed based on both a long-term static metric (Betweenness-centrality) and two short-term dynamic metrics (the number of Pending Interests and the number of data chunks). After smoothing out the short-term parameters using EWMA, a data fusion technique based on PCA is applied to extract the principle factors of this data set. Using these factors, the quota for the cache size of each router is obtained. A typical scenario based on Abilene core topology is implemented in NS-3 simulator to evaluate the performance of the proposed method. Evaluation results show that the proposed method does not just increase hit ratio of the routers, but it also has the advantage of exploiting the information of graph-based centrality metric without considerable computation in nodes and message-passing between them. Future efforts for extending this study involve applying mobility in NDN for real-time applications \cite{31}; in other words, we aim to incorporate other parameters such as influence of mobile \textit{Consumer} devices on cache size allocation scheme of different routers. Then by exploiting another matrix decomposition data-fusion approach \cite{32}, importance of each router is computed.

\Acknowledgements{The authors would like to thank Fereshte Dehghani and Elahe Hadi for their suggestions and feedbacks.}

\end{document}